\RequirePackage{fix-cm}
\documentclass[smallextended]{svjour3}       
\smartqed  
\usepackage{graphicx}
\usepackage{amsmath}
\usepackage{amssymb}

\newcommand{\alo}{Al$_2$O$_3$/V}
\begin{document}

\title{A Soft X-ray Beam-splitting Multilayer Optic for the NASA GEMS Bragg Reflection Polarimeter
}

\titlerunning{GEMS BRP Reflector}        

\author{Ryan Allured         \and
        M\'{o}nica Fern\'{a}ndez-Perea \and
        Regina Soufli \and
        Jennifer B. Alameda \and
        Michael J. Pivovaroff \and
        Eric M. Gullikson \and
        Philip Kaaret
}


\institute{Ryan Allured \and Philip Kaaret \at
              University of Iowa, Iowa City, IA, USA \\
              \email{ryan-allured@uiowa.edu}           
           \and
           M\'{o}nica Fern\'{a}ndez-Perea \and
        Regina Soufli \and
        Jennifer B. Alameda \and
        Michael J. Pivovaroff \at
              Lawrence Livermore National Laboratory, Livermore, CA, USA
           \and
           Eric M. Gullikson \at
           Lawrence Berkeley National Laboratory, Berkeley, CA, USA
}

\date{Received: date / Accepted: date}

\maketitle

\begin{abstract}
A soft X-ray, beam-splitting, multilayer optic has been developed for the Bragg Reflection Polarimeter (BRP) on the NASA Gravity and Extreme Magnetism Small Explorer Mission (GEMS). The optic is designed to reflect 0.5 keV X-rays through a $90^\circ$ angle to the BRP detector, and transmit 2--10 keV X-rays to the primary polarimeter. The transmission requirement prevents the use of a thick substrate, so a 2 $\mu$m thick polyimide membrane was used. Atomic force microscopy has shown the membrane to possess high spatial frequency roughness less than 0.2 nm rms, permitting adequate X-ray reflectance. A multilayer thin film was especially developed and deposited via magnetron sputtering with reflectance and transmission properties that satisfy the BRP requirements and with near-zero stress. Reflectance and transmission measurements of BRP prototype elements closely match theoretical predictions, both before and after rigorous environmental testing.
\end{abstract}

\section{Introduction}
\label{sec:Intro}  

\subsection{BRP Overview}

The Bragg Reflection Polarimeter (BRP) was the student experiment on the NASA Gravity and Extreme Magnetism Small Explorer (GEMS) mission \cite{Swank10}.  The experiment existed under the NASA Education and Public Outreach program, and its primary purpose was to train students in the field of space exploration.  The scientific goal was to examine the geometry of spacetime surrounding accreting black holes using X-ray polarization.  A team of students at the University of Iowa (UI) was tasked with the design, construction, testing, integration, and operation of a soft X-ray polarimeter under the guidance of UI faculty and NASA personnel.  The BRP instrument and the GEMS mission passed the Preliminary Design Review at Goddard Space Flight Center (GSFC). However, the mission was subsequently canceled for budgetary reasons.  At the time of cancellation, the team was developing flight hardware and preparing for the Critical Design Review.  This paper reports on the development of the multilayer reflector for this instrument, following an earlier conference proceedings publication on this topic \cite{SPIEReflector}.

\begin{figure}
\includegraphics[width=\columnwidth]{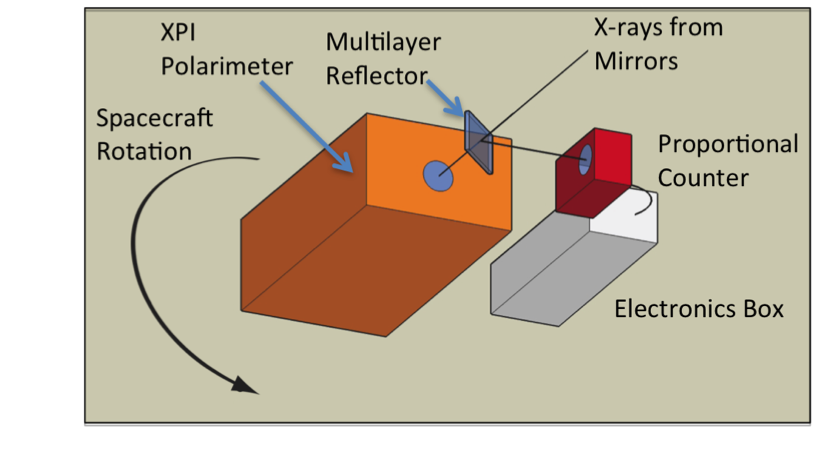}
\caption{\label{fig:BRPconcept}A conceptual diagram of the operation of the GEMS BRP.  The multilayer reflector is placed in the optical path of the main experiment.  The proportional counter monitors the X-ray count rate that is modulated as a function of spacecraft roll angle, depending on the polarization of the incoming beam.}
\end{figure} 

The basic instrument concept is shown in Fig.\ \ref{fig:BRPconcept}.  Glancing incidence Wolter Type I mirrors focus a cone of light toward the primary experiment (XPI: X-ray Polarimeter Instrument \cite{Black10}).  The multilayer reflector is placed in this optical path at a nominal $45^\circ$ incidence angle, reflecting a narrow energy band centered at about 0.5 keV to the multiwire proportional counter (MWPC).  Due to the polarization dependence of reflection, the reflected intensity will vary based on the polarization of the incoming beam and the roll angle of the spacecraft.  As the spacecraft rotates about the optical axis, the MWPC monitors the count rate to obtain a polarization measurement of an astrophysical source.

Three main components make up the BRP: the multilayer reflector, the MWPC, and the electronics box.  The reflector consists of a polyimide membrane substrate mounted on a metallic frame and coated with an ultra-short-period multilayer coating.  The MWPC is loosely based on the design of the ROSAT PSPC \cite{Briel86}, and possesses discrete regions for X-ray detection and charged particle background rejection.  The electronics box supplies high voltage to the MWPC, shapes and digitizes MWPC pulses, and communicates with the spacecraft.  
 
\subsection{Reflector Requirements}

A common figure of merit for a polarimeter is the minimum detectable polarization (MDP), defined as the 99\% confidence limit that a polarization amplitude is not measured by chance.  For a source rate $S$, background rate $B$, and observation time $T$, the fractional MDP is given by \begin{equation} {\rm MDP}=\frac{4.29}{\mu S} \left( \frac{S+B}{T} \right) ^{1/2}, \label{eq:MDP}\end{equation} where the modulation factor $\mu$ is the polarimeter response to a 100\% polarized source \cite{Weisskopf09}.  In the X-ray band, Brewster's angle for 0\% reflection of p-polarized light is $\sim 45^\circ$, and so for a Bragg polarimeter operating at $45^\circ$ incidence $\mu \simeq1$.

BRP MDP estimates were made using the energy spectrum of the Crab Nebula, the only X-ray source to have its polarization reliably measured (15.7\% and 18.3\% at 2.6 keV and 5.2 keV, respectively \cite{Weisskopf76}) in the past.  Theoretical reflectance curves (see \S \ref{sec:modeling}) for various multilayer designs were convolved with the Crab energy spectrum \cite{Zombeck07} to predict the source rate $S$, and a rough estimate for the detector background rate $B$ (0.006 cts/sec) was obtained using the background rates, scaled for the detector size, on the OSO-8 Wisconsin Experiment \cite{Bunner78}.  The Crab observation time was fixed to that in the GEMS Observing Plan ($3.8\times10^5$ sec).  The multilayer integrated reflectance at a fixed incidence of $45^\circ$ was used to quantify the reflector sensitivity requirement, which was set at 0.08 eV based on the reflectance curve of the multilayer design with the highest permissible predicted MDP (10\%).  The rms slope of the reflector surface, calculated at spatial wavelengths between 1 mm and the full aperture, was required to be $< 5$ arcmin to ensure the reflected photons pass through the 11.2 mm diameter detector window.  A larger window with sufficient mechanical strength and soft X-ray transmission was not commercially available.

In order to not reduce the XPI sensitivity, the reflector must have high transmission in the 2--10 keV band.  Transmission curves were calculated and convolved with the XPI effective area to predict the reduction in sensitivity due to the BRP.  In this calculation, the polyimide substrate thickness was fixed at 2 $\mu$m and the number of bilayers was varied for each candidate multilayer material pair.  A 70\% transmission requirement was set at 2.7 keV using the calculated transmission curve that produced the maximum allowable reduction in XPI sensitivity.  This energy is at the low end of the XPI energy band, where the transmission is expected to be the worst.  Also, it is straightforward to measure transmission at this energy using an X-ray tube and a rhodium target---the full transmission curve can then be extrapolated from this measurement.  A clear aperture requirement was set at 32 mm so that the metallic frame would not occult the light cone seen by the XPI window.

\section{Substrate Selection and Characterization}

A thin film substrate for the multilayer coating was necessary to satisfy the transmission requirement.  Approximately $2~\mu$m thick polyimide samples were obtained from Luxel Corp.\ for surface quality and X-ray transmission measurements.  The polyimide membranes were mounted on aluminum rings with a 28.6 mm inner diameter.

\subsection{Surface Roughness}

The ultimate reflectance achieved by a multilayer structure is reduced from its theoretical maximum due to imperfections in the individual interfaces.  Sources of imperfection include surface roughness, interdiffusion of adjacent layers, and even the size of the atoms themselves.  Typically, substrate surface roughness propagates through the layers of a multilayer.  A rule of thumb is to choose a substrate with rms roughness $< 0.1d$, where $d$ is the period or bilayer thickness of the multilayer \cite{Spiller94}.  The Bragg equation ($m\lambda=2d{\rm sin}\theta$) leads to a bilayer thickness of $\sim1.7$ nm for a peak reflectance at $\sim0.5$ keV.  Therefore, we sought a thin film substrate with roughness $\lesssim 2$ \AA~rms.

\begin{figure}
\includegraphics[width=\columnwidth]{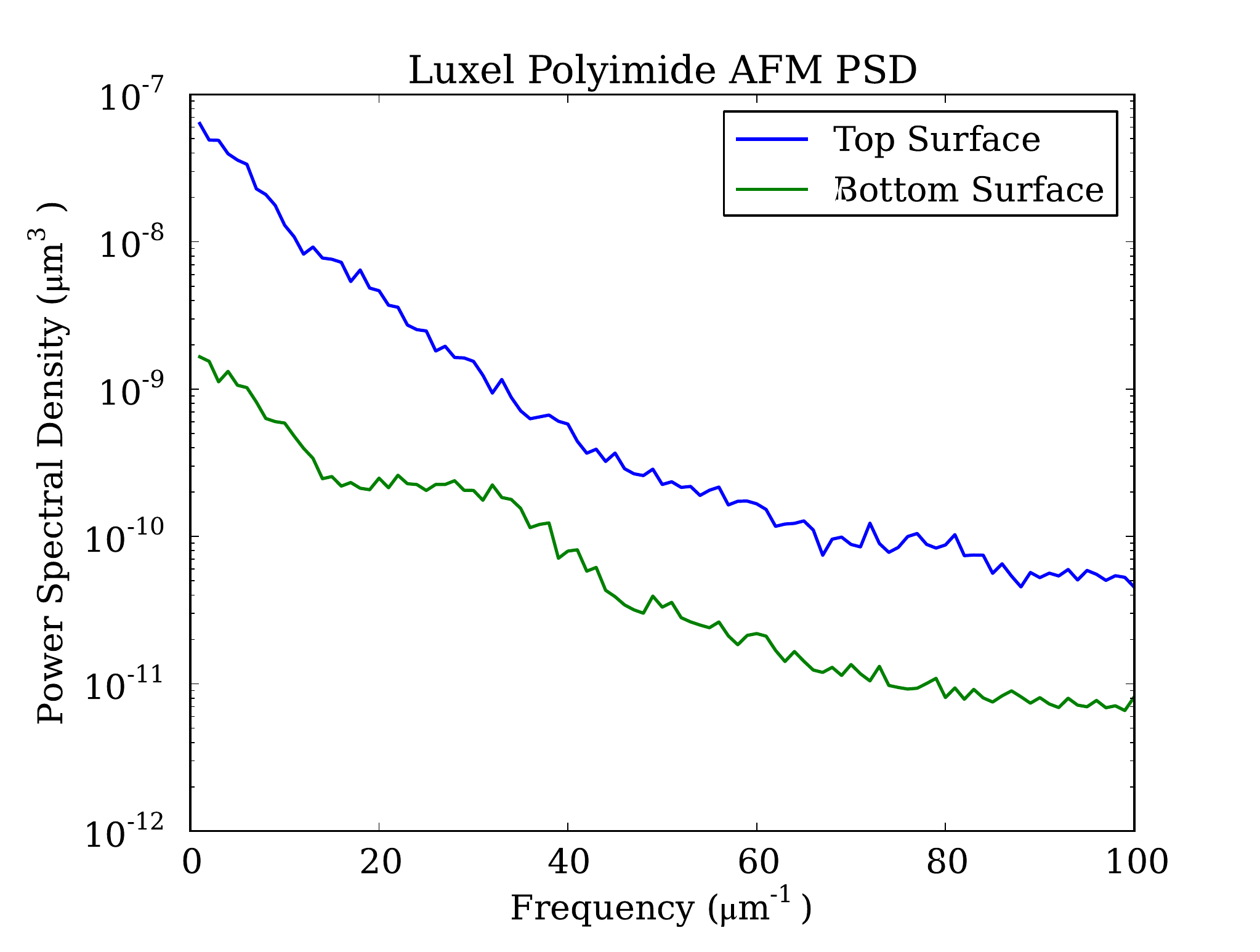}
\caption{\label{fig:LuxelPSDs}One dimensional PSDs of the Luxel polyimide sample, calculated using 1 $\mu\rm{m}^2$ AFM measurements and averaged over 512 scans.  The blue line shows the PSD of the surface of the film exposed to vacuum (top) during substrate fabrication, whereas the green line shows the PSD of the surface making contact with the Si wafer (bottom).}
\end{figure} 

The spatial frequency range of surface imperfections that degrade soft X-ray reflectance is approximately 1--100 $\mu {\rm m}^{-1}$ \cite{Windt94}.  Atomic force microscopy (AFM) is the only metrology technique able to access this frequency range.  Using the Asylum Research MFP-3D located at the UI Central Microscopy Research Facility, profiles of both surfaces of a polyimide sample were mapped out using 1 $\mu\rm{m}^2$ scans with 512 by 512 points (1 scan per surface).  One dimensional power spectral densities (PSDs) were calculated and are shown in Fig.\ \ref{fig:LuxelPSDs}.  The bottom surface refers to the side of the film making contact with the Si wafer during substrate fabrication.  The roughnesses obtained by integrating the PSDs in Fig.\ \ref{fig:LuxelPSDs} over the 1--100 $\mu$m$^{-1}$ range are 6.76 \AA ~and 1.34 \AA~rms for the top and bottom surfaces, respectively.  The roughness of the bottom surface closely matches the ultra smooth Si wafer onto which it was deposited, which has a typical 1 \AA ~rms roughness in the same spatial frequency range.

\begin{figure}[ht]
\begin{minipage}[t]{0.5\columnwidth}
\centering
\includegraphics[width=\textwidth]{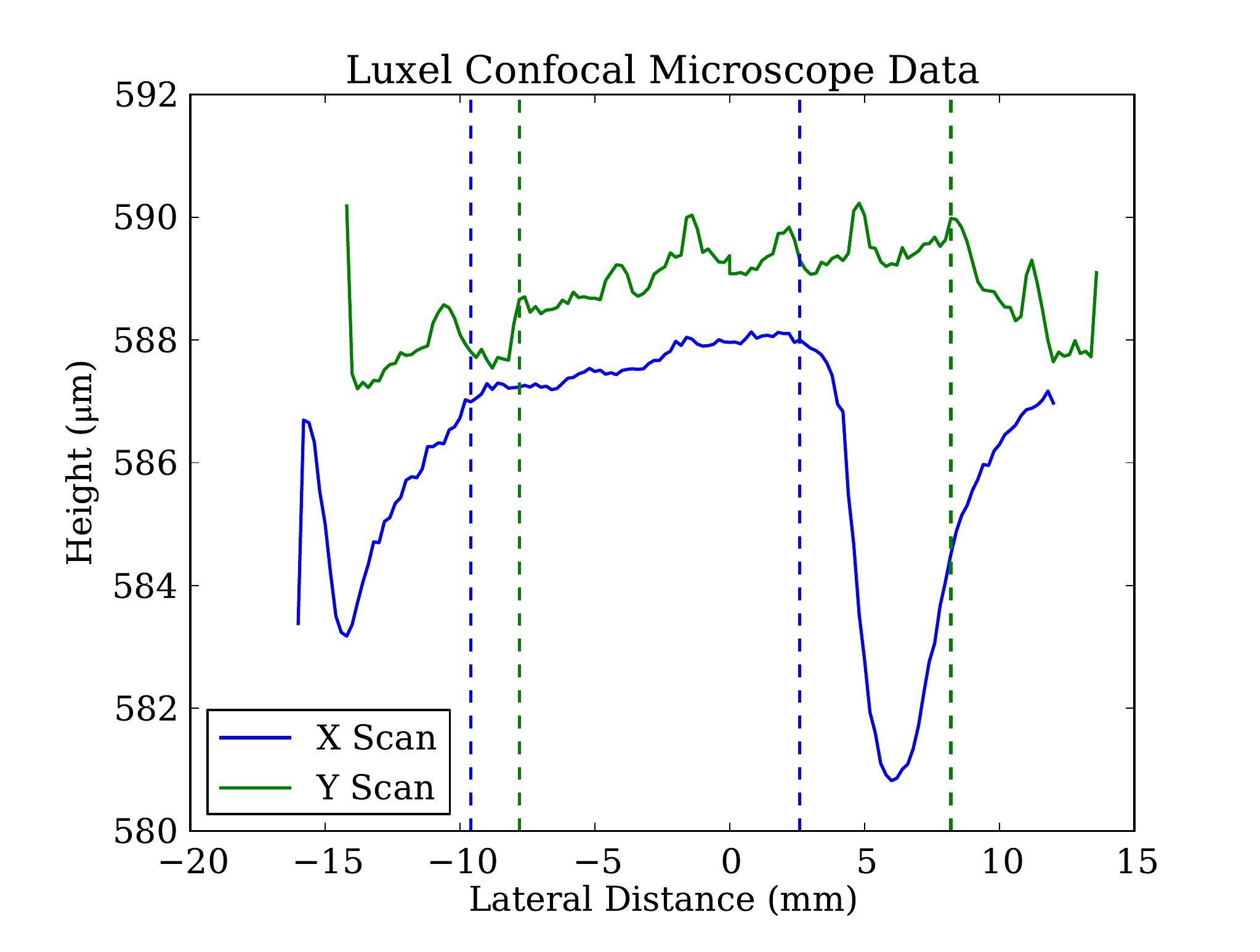}
\caption{The confocal microscope scans, provided by Luxel Corp., of the bottom surface of the polyimide membrane. Vertical dashed lines indicate the limits chosen for PSD calculation.}
\label{fig:LuxelScan}
\end{minipage}
\hspace{0.5cm}
\begin{minipage}[t]{0.5\columnwidth}
\centering
\includegraphics[width=\textwidth]{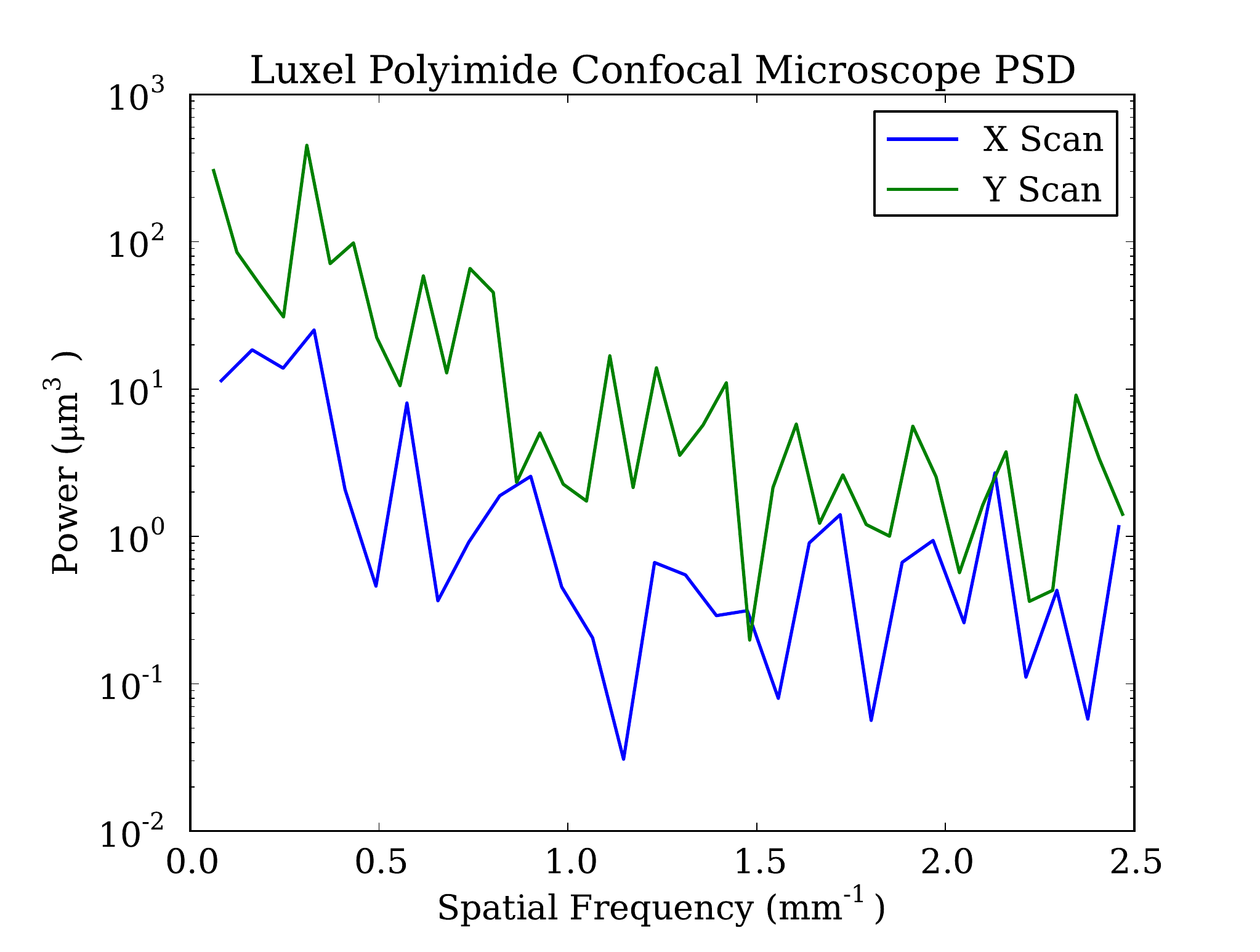}
\caption{One dimensional PSDs of the bottom surface of the Luxel polyimide, calculated using orthogonal confocal microscope scans across the surface. Note that the spatial frequencies shown here do not overlap with those of Fig.\ \ref{fig:LuxelPSDs}}
\label{fig:ConfocalPSD}
\end{minipage}
\end{figure}
  
\subsection{Slope Error}

Variation of surface heights in spatial frequencies $\lesssim 1~ {\rm mm}^{-1}$ has little effect on reflectance throughput, and instead determines the quality of the reflected image.  A reflector with significant slope error (i.e. waviness) will act to smear the point spread function of the image at the detector window (11.2 mm diameter), resulting in a lower encircled energy fraction.  Combined with the pointing wobble of the spacecraft, this effect will increase the uncertainties in a polarization measurement.  To limit this impact, slope error is required to be $< 5$ arcmin in the spatial period range of 1 mm to the full aperture of the reflector.  The clear aperture of the BRP reflector for the flight instrument is required to be $>32$ mm. For testing purposes, the polyimide substrates discussed in this manuscript have a diameter of 28.6 mm.

Using Luxel supplied confocal microscope scans (Fig.\ \ref{fig:LuxelScan}) of the polyimide substrate, PSDs were calculated over the spatial frequency range of $\sim 0.1$--$2.5 {\rm ~mm}^{-1}$ (Fig.\ \ref{fig:ConfocalPSD}).  As discussed later in \S\ref{sec:ripples}, ripples such as the one observed in the right side of the x scan in Fig.\ \ref{fig:LuxelScan} are expected to nearly disappear under the environment of a low earth orbit. They would also be less pronounced if a material with a smaller coefficient of thermal expansion was used for the membrane support ring. Therefore, we restricted the slope error calculation to ripple-free slices of the membrane surface as shown in Fig.\ \ref{fig:LuxelScan}. The slope error was quantified by integrating the second moment of the resultant PSDs, and was 3.47 arcmin and 1.34 arcmin for the x and y scans, respectively.  Both of these numbers are well within the 5 arcmin requirement.

\subsection{X-ray Transmission}

\begin{figure}[ht]
\begin{minipage}[t]{0.5\columnwidth}
\centering
\includegraphics[width=\textwidth]{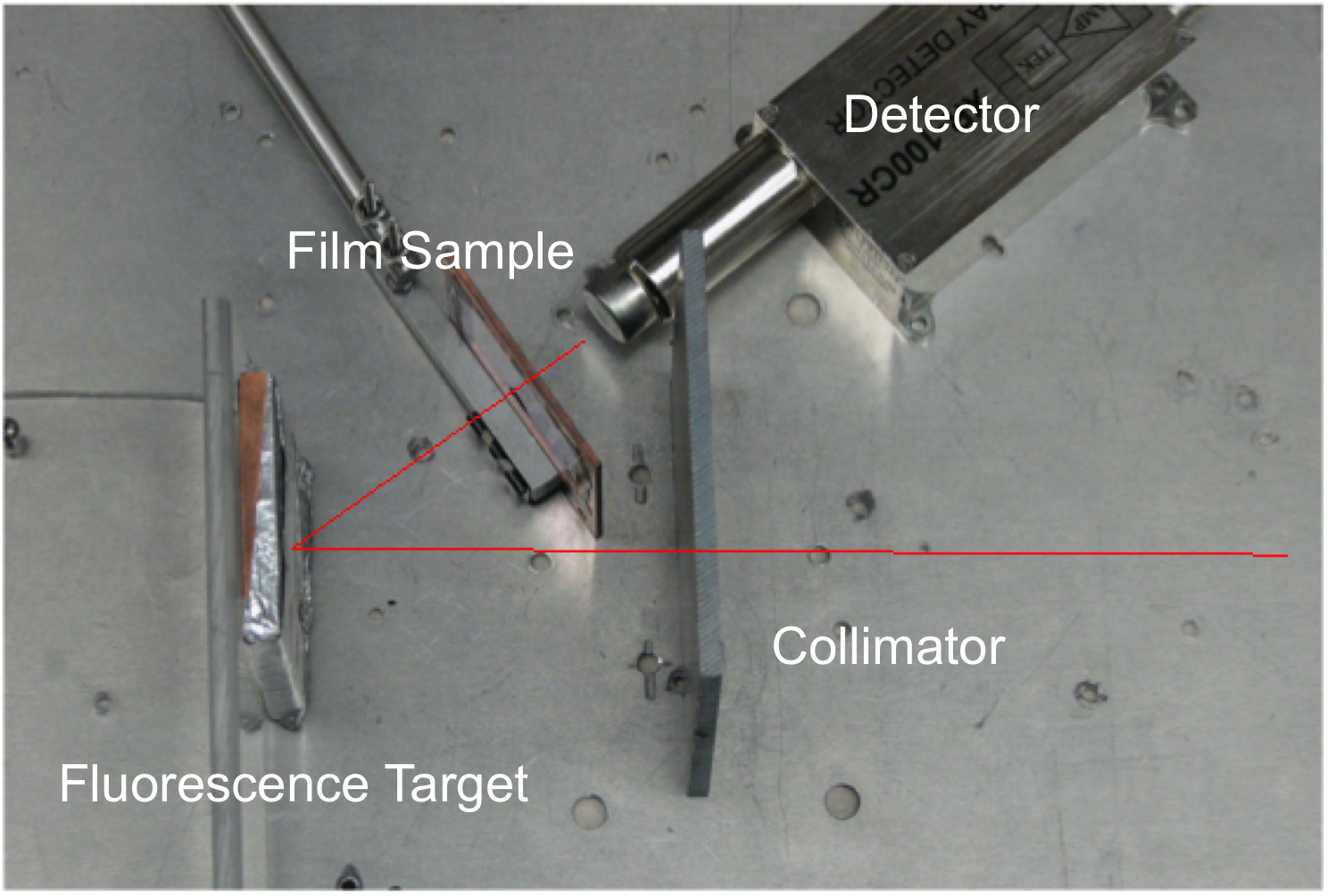}
\caption{The setup used to measure the 1.5 keV transmission of the Luxel polyimide.  The X-rays enter from the right, pass through a stainless steel collimator, and irradiate the copper target wrapped with aluminum foil.  The fluorescence X-rays pass through the polyimide sample before entering the Amptek XR-100CR silicon detector.  The red line indicates the path of the entering X-rays and the fluorescence X-rays entering the detector.  The film sample shown here is mylar investigated prior to the Luxel polyimide.}
\label{fig:TransSetup}
\end{minipage}
\hspace{0.5cm}
\begin{minipage}[t]{0.5\columnwidth}
\centering
\includegraphics[width=\textwidth]{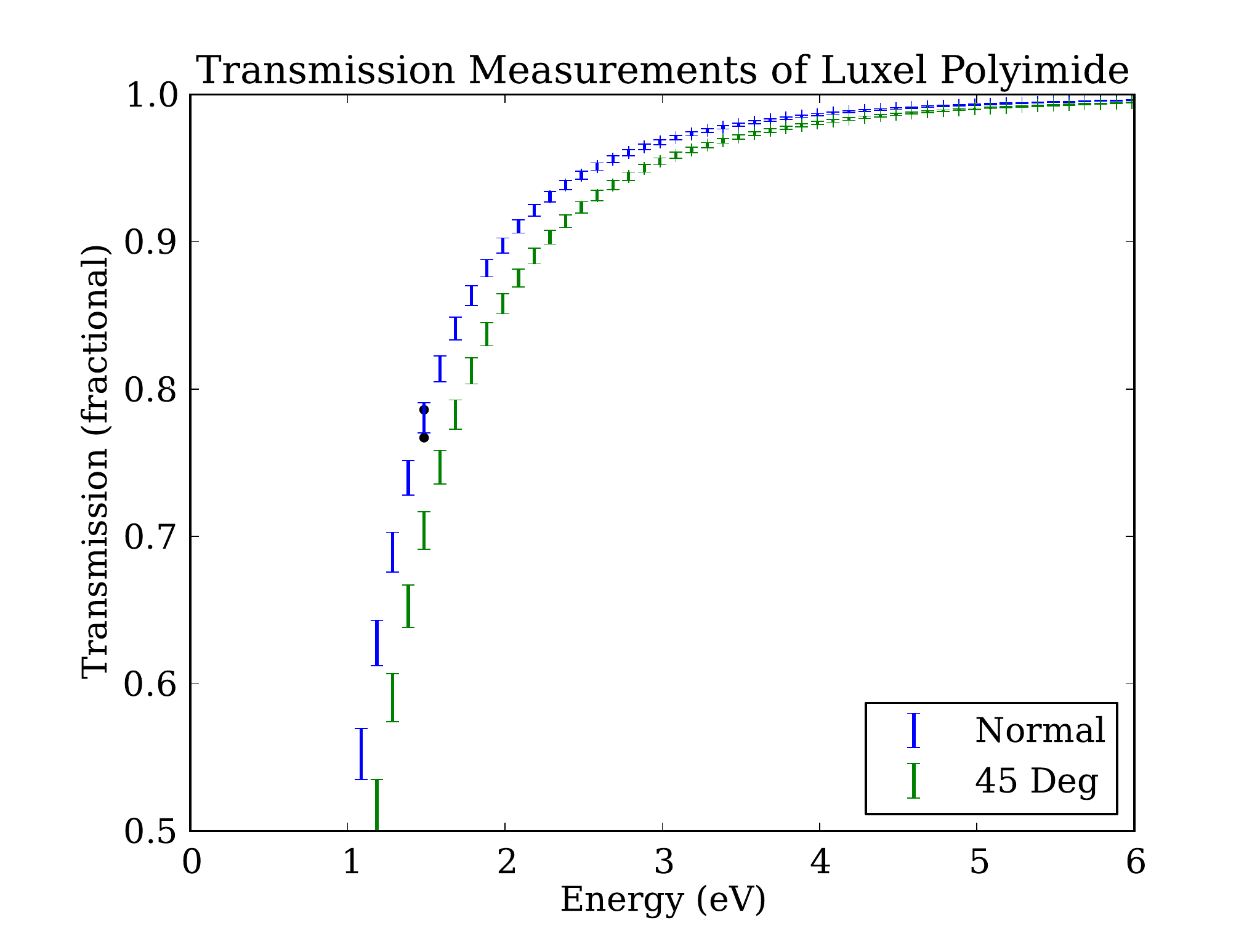}
\caption{Results of the substrate transmission measurements, indicated by the black points, compared to a model calculated with data obtained from the CXRO website \cite{Henke93}. The error bars were obtained assuming a $\pm 0.1~\mu$m maximum variation in the polyimide thickness.}
\label{fig:SubTrans}
\end{minipage}
\end{figure}

The X-ray transmission of the substrate alone was tested at normal incidence prior to reflector development in order to ensure that the 70\% transmission requirement for the multilayer-coated substrate could be met at 2.7 keV.  The experimental setup is depicted in Fig.\ \ref{fig:TransSetup}.  A collimated beam generated by an Oxford X-ray tube enters the chamber and impacts a target consisting of a copper plate covered with aluminum foil.  The tube's rhodium anode is varied between trials over 20--25 kV in order to produce a strong bremsstrahlung continuum for fluorescing aluminum and copper.  An Amptek XR-100CR silicon detector is positioned off-axis to monitor the resulting 8 keV copper and 1.5 keV aluminum lines.  The polyimide sample is moved in and out of the path between the target and the detector with a linear vacuum feedthrough.  The aluminum line count rate with the sample in the fluorescence path divided by that with the sample out of the fluorescence path is taken as the transmission.  The copper line, which has $> 99.8\%$ transmission through the polyimide, is used to normalize spectra from run to run due to the varying X-ray tube flux.

The results of the measurement are shown in Fig.\ \ref{fig:SubTrans}.  The black points indicate the 76.7\% and 78.6\% measured transmission values for two samples of Luxel polyimide.  Theoretical transmission values were calculated over the GEMS energy range using the Center for X-ray Optics online database\cite{Henke93}.  The measurements are consistent with the theoretical transmission values within the $\pm 0.1 ~\mu{\rm m}$ uncertainty in the film thickness, lending confidence to the ability to predict substrate transmission using tabulated optical constants.  The extrapolated transmission at 2.7 keV and a $45^\circ$ incidence angle is 94.0\%, deemed sufficient for the reflector substrate.

\section{Multilayer Modeling and Material Pair Selection}
\label{sec:modeling}

Multilayer modeling using the IMD software package was used to select the multilayer materials and optimal parameters \cite{Windt98}.  The primary considerations were the theoretically achievable reflectance and transmission, systematic errors induced by the spacecraft pointing wobble, and the potential stress put on the substrate by the multilayer coating, which should be minimized to ensure polyimide membrane survival.  The candidate material pairs were WC/SiC and \alo.  WC/SiC was specifically designed to have near-zero compressive stress, and \alo~has high theoretical reflectance below the vanadium L$_{2,3}$ edge at 512 eV.

\subsection{Weighted Reflectance Curves}

\begin{figure}
   \begin{center}
   \begin{tabular}{c}
   \includegraphics[width=\columnwidth]{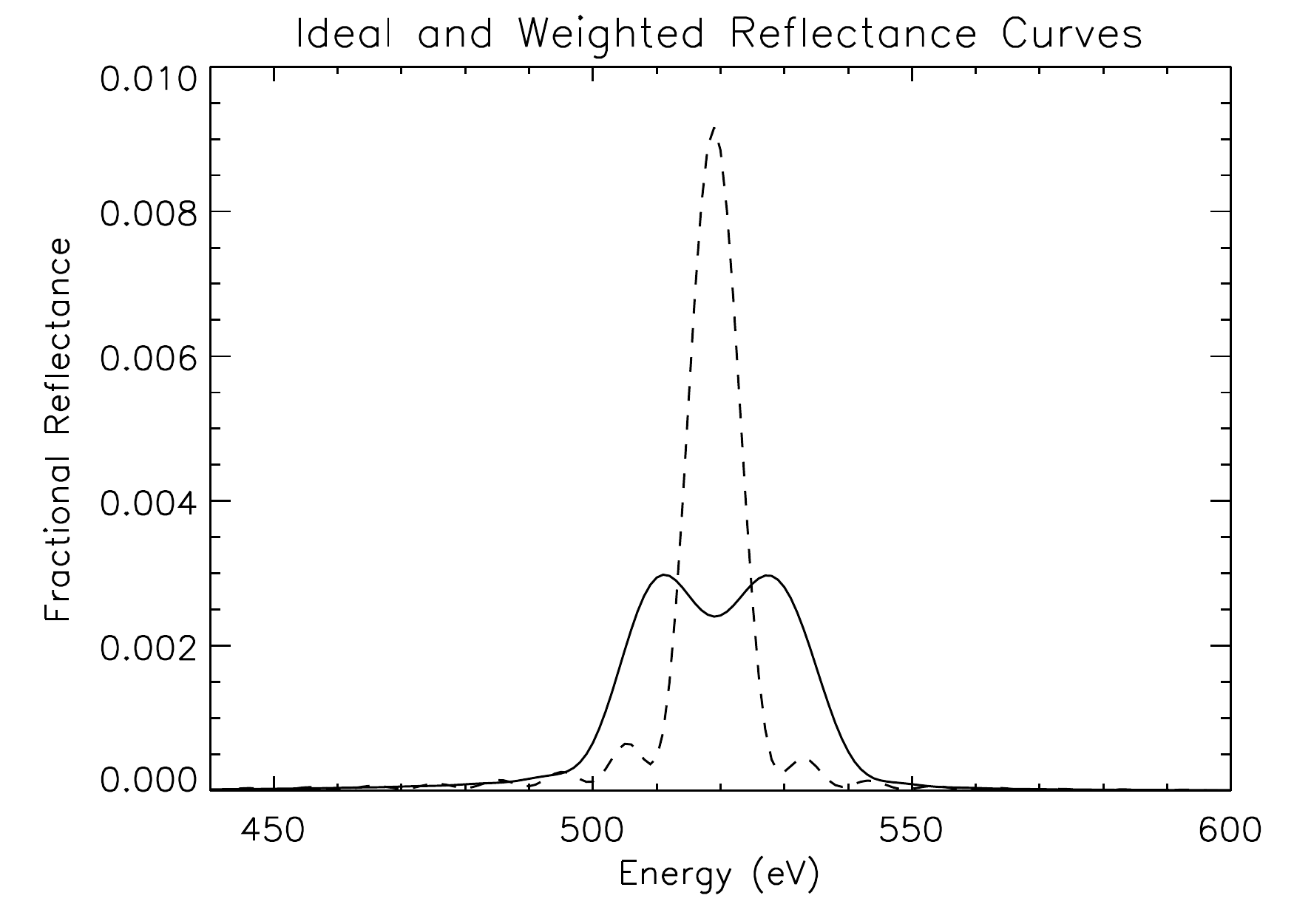}
   \end{tabular}
   \end{center}
   \caption[example] 
   { \label{fig:BlendedCurve} 
An ideal WC/SiC reflectance curve calculated for a nominal $45^\circ$ incidence angle is shown by the dashed line. A reflectance curve calculated using weights from the simulated GEMS incidence angle distribution is shown by the solid line. Reflectance values in this plot were calculated with arbitrary surface roughness and are therefore not representative of those predicted for the flight reflector.}
   \end{figure}

Although the BRP reflector's nominal incidence angle is $45^\circ$, the converging X-ray cone hitting the reflector will have an inherent spread in incidence angles of approximately $\pm2^\circ$, dictated by the geometry of the Wolter I mirrors.  An approximate form of this incidence angle distribution was determined using a simplified 2D model of the mirrors.  This angular distribution was used to weight reflectance curves calculated for incidence angles covering the $\pm2^\circ$ range, which were then summed together to arrive at the true reflector response versus energy.  Fig.\ \ref{fig:BlendedCurve} shows two WC/SiC reflectance curves.  The sharply peaked curve represents the predicted response for a nominal $45^\circ$ incidence angle.  The smooth, broadly peaked curve is the result from the weighted averaging of reflectance curves over the $45\pm2^\circ$ range.  While the peak reflectance of the weighted curve is significantly lower than in the ideal case, the broader energy response acts to preserve the integrated reflectance.  Because the optical constants of vanadium are rapidly varying in our energy range, integrated reflectivity is not preserved under this transformation for \alo~multilayers.

\subsection{MDP Optimization}

For each multilayer material pair candidate (\alo~and WC/SiC), IMD was used to scan through coating parameters.  Mass densities between 90--100\% of bulk, typical for sputtered thin films, were used as input parameters for the constituent layer materials in each multilayer coating.  Variable parameters included the ratio ($\Gamma$) of high-Z layer thickness to bilayer period and the energy of peak reflectance (mainly dictated by the bilayer thickness $d$ and incidence angle).  For the WC/SiC case, $\Gamma$ was fixed to 0.5 in order to achieve near-zero compressive stress (see \S \ref{sec:MultilayerDesign}).  The nominal incidence angle for the WC/SiC coatings was varied as an academic exercise, but only 45$^\circ$ was considered for flight due to geometrical mounting constraints.  The optimal multilayer coating parameters were determined on the basis of the design with the lowest theoretical MDP.

For a given set of coating parameters, the maximum number of bilayers $N_{max}$ was determined by calculating the multilayer transmission of 2.7 keV photons at a $45^\circ$ incidence angle, and taking the highest number of layers for which the transmission, including the effect of the substrate, was $>75\%$ (for margin on the 70\% requirement).  The weighted reflectance curve for unpolarized light was then calculated for $N_{max}$ bilayers.  This curve was convolved with the Crab nebula energy spectrum \cite{Zombeck07} between 350 and 650 eV.  The integral of this curve was multiplied by the effective area of the GEMS mirrors at 500 eV (207 cm$^2$), the transmission of the mirror thermal shields at 500 eV (39.69\%), and the transmission of the BRP detector window at 500 eV ($\sim 47\%$) to calculate an average count rate.  The quantum efficiency of the detector ($\sim1$) and the encircled energy fraction ($\sim 100\%$) were ignored.  Finally, the count rate was inserted into Eq.\ \ref{eq:MDP} to arrive at an MDP.

\begin{figure}[ht]
\begin{minipage}[t]{0.5\columnwidth}
\centering
\includegraphics[width=\textwidth]{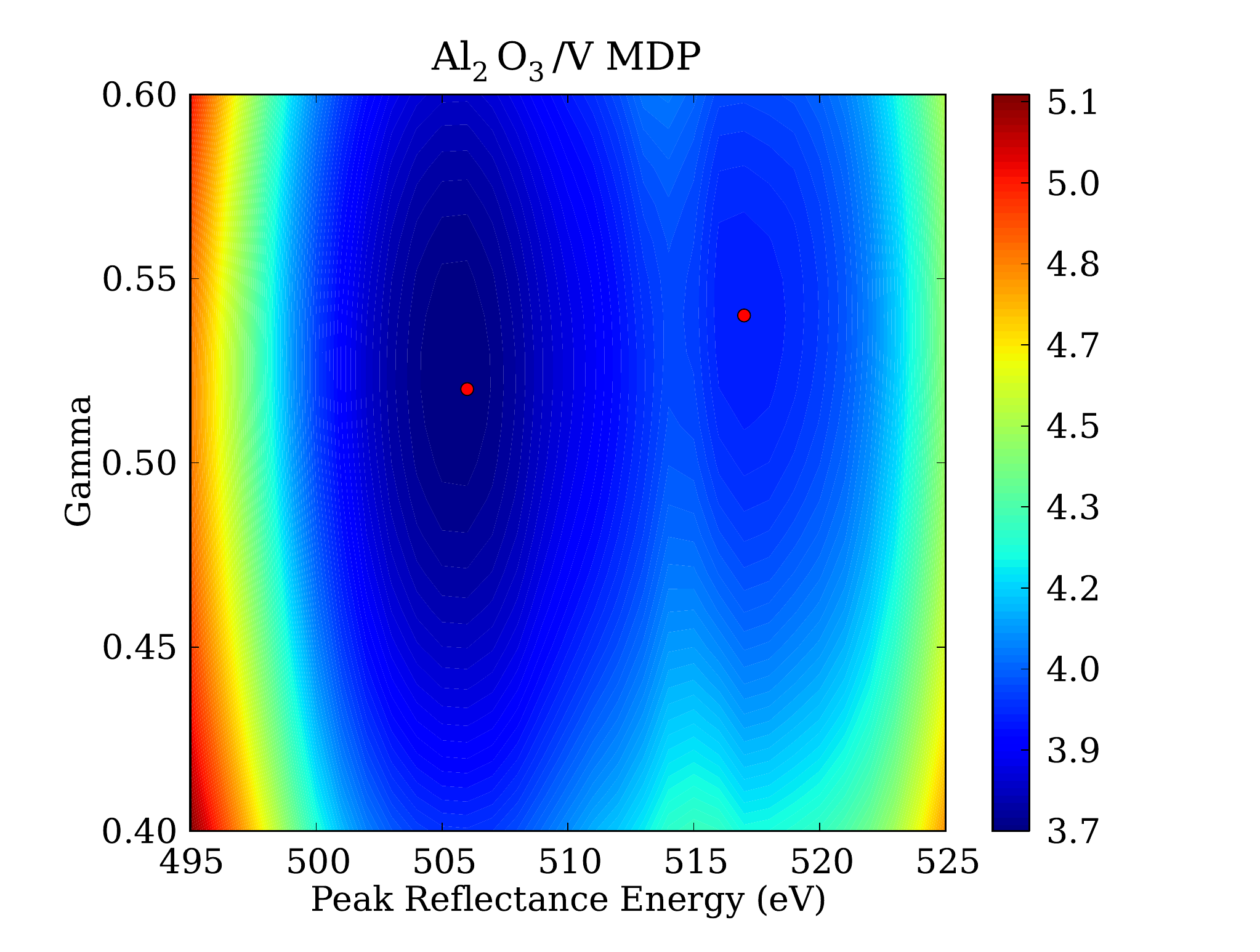}
\caption{\alo~ parameter optimization results.  Minimum detectable polarization is plotted as a function of peak reflectance energy and the ratio $\Gamma$ of V thickness to bilayer thickness.  The two local minima at 506 eV, 0.52 and 517 eV, 0.54 are marked with red dots.}
\label{fig:AlOpt}
\end{minipage}
\hspace{0.5cm}
\begin{minipage}[t]{0.5\columnwidth}
\centering
\includegraphics[width=\textwidth]{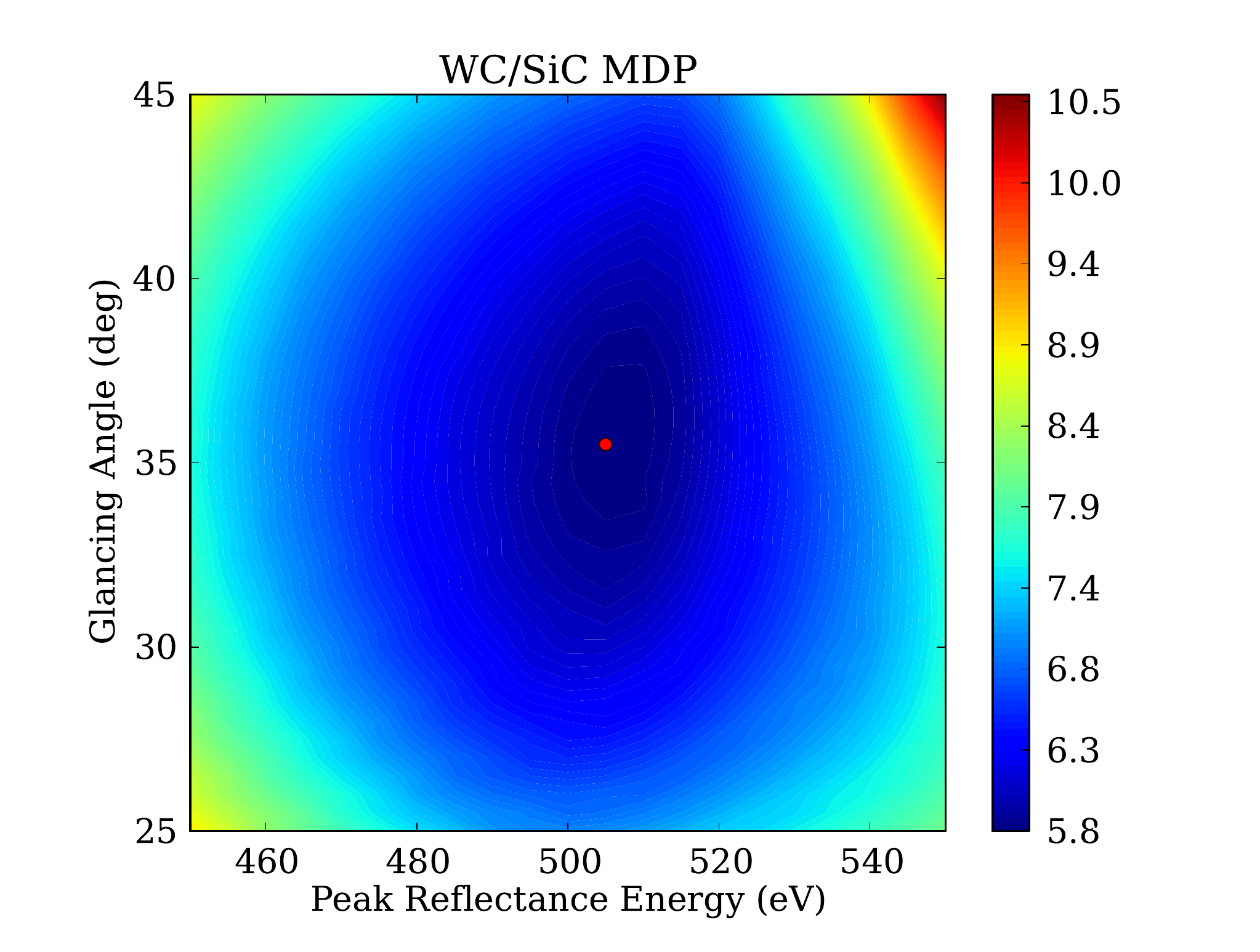}
\caption{WC/SiC parameter optimization results.  Minimum detectable polarization is plotted as a function of peak reflectance energy and the nominal incidence angle. The minimum at 506 eV, 33$^\circ$ is marked with a red dot.}
\label{fig:WCOpt}
\end{minipage}
\end{figure}

The results of the \alo~ optimization (varying $\Gamma$ and peak reflectance energy) are shown in Fig.\ \ref{fig:AlOpt}.  Two local minima can be seen in the contours.  The global minimum is at a peak reflectance energy of 506 eV and a $\Gamma$ of 0.52.  The low MDP at this energy is attributed to high theoretical peak reflectance just below the vanadium L$_{2,3}$ edge at 512 eV.  The other minimum is at an energy of 517 eV and a $\Gamma$ of 0.54.  At this peak energy, the \alo~ weighted reflectance curve lies just below the oxygen absorption edge of the interstellar medium, where the flux from the Crab is the highest within our scanning region.

The WC/SiC optimization results are shown in Fig.\ \ref{fig:WCOpt}.  Note that here the incidence angle is used as the second optimization parameter instead of $\Gamma$, which is fixed at 0.5. Because the optical constants of these materials are smooth in our energy range, the MDP is a slowly varying function.  With the incidence angle fixed to $45^\circ$, the optimal peak reflectance energy is 510 eV.  The global minimum is at a peak reflectance energy of 505 eV and a $35.5^\circ$ incidence angle.  As the incidence angle becomes more glancing, the $\pm2^\circ$ angular spread creates a wider energy response due to the sinusoidal angular dependence of the Bragg equation.  This causes the optimal peak reflectance energy to be pushed toward lower energies in order to avoid the oxygen absorption edge as the incidence angle is reduced.

At the 505 eV, $35.5^\circ$ global minimum, the predicted modulation factor ($\mu$) is reduced to 0.7 but the predicted count rate is $\sim 2$ times greater than that at 510 eV, $45^\circ$.  While a lower modulation factor should increase the expected MDP, the higher reflectance achieved at more grazing angles increases the source count rate and leads to a lower MDP.  The optimal angle for Bragg reflection polarimetry is found by balancing a higher source rate with a reduced modulation factor at more glancing incidence angles, and depends on the specific multilayer/crystal design in question.  An additional consideration in our case is the increasingly limited number of bilayers, and hence reflectance, due to reduced transmission at more glancing angles.

   \begin{figure}
   \begin{center}
   \begin{tabular}{c}
   \includegraphics[width=\columnwidth]{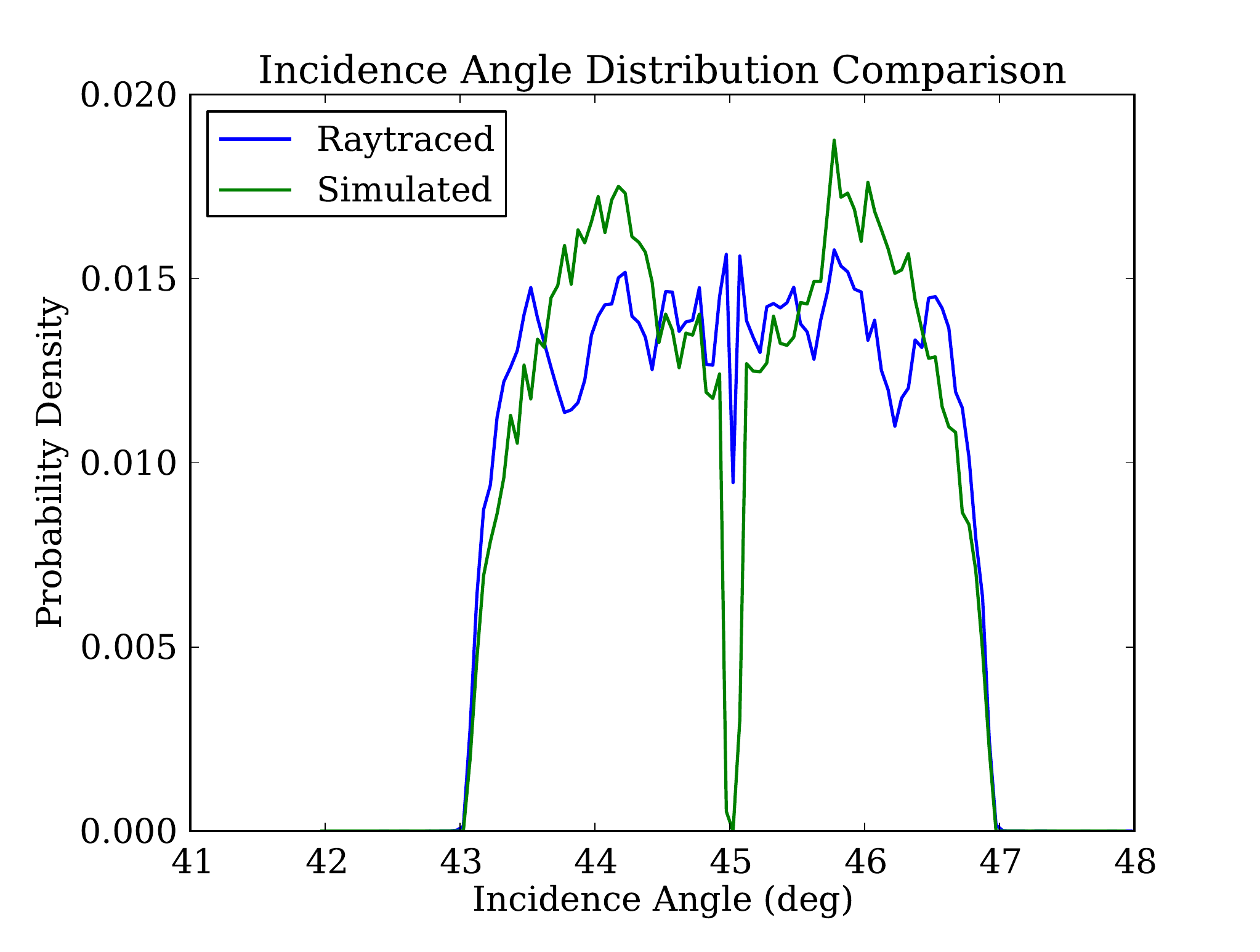}
   \end{tabular}
   \end{center}
   \caption[example] 
   { \label{fig:IncDistComp} 
The GEMS incidence angle distribution obtained from a simple 2D mirror model is shown in green, and that obtained from the GEMS raytracing code is shown in blue.}
   \end{figure} 
   
The above modeling made use of reflectance curves weighted by the $\pm2^\circ$ GEMS incidence angle distribution.  The form of the distribution was obtained by taking random points from a 2D model of the GEMS mirrors and tracing them to the focal point.  Later, a more realistic GEMS incidence angle distribution was obtained from the GEMS raytracing code.  The two distributions, shown in Fig.\ \ref{fig:IncDistComp}, are extremely similar.  Substituting the distribution from GEMS raytracing into the MDP optimization code resulted in less than 1\% (relative) difference in the contour plots shown in Figs.\ \ref{fig:AlOpt} and \ref{fig:WCOpt}.  

\subsection{Systematic Error}

In a perfectly aligned and pointed telescope, the optimal parameters found in the previous section would be chosen for fabrication (i.e. an \alo~multilayer with peak reflectance energy at 506 eV and $\Gamma=0.52$).  However, false polarization signals due to telescope misalignment and spacecraft pointing wobble must be taken into account.  For example, a fixed telescope misalignment could cause the beam to wobble about the nominal beam axis.  Because reflectance is dependent on incidence angle, this could cause a modulation in count rate unrelated to the inherent polarization of the incident beam.  A rough simulation was carried out for the purposes of multilayer material pair selection.

The predicted GEMS pointing accuracy at the time of material pair selection was $\sim 2$ arcmin.  The worst case false polarization induced by misalignments assumed a fixed 2 arcmin pointing error.  Count rates were calculated for $\pm2$ arcmin from the nominal $45^\circ$ incidence angle, and induced polarization was calculated as $(R_{max}-R_{min})/(R_{max}+R_{min})$.  This calculation was repeated for nominal incidence angles of $44.7^\circ$ and $45.3^\circ$ to account for potential reflector misalignment.

\begin{figure}[ht]
\begin{minipage}[t]{0.5\columnwidth}
\centering
\includegraphics[width=\textwidth]{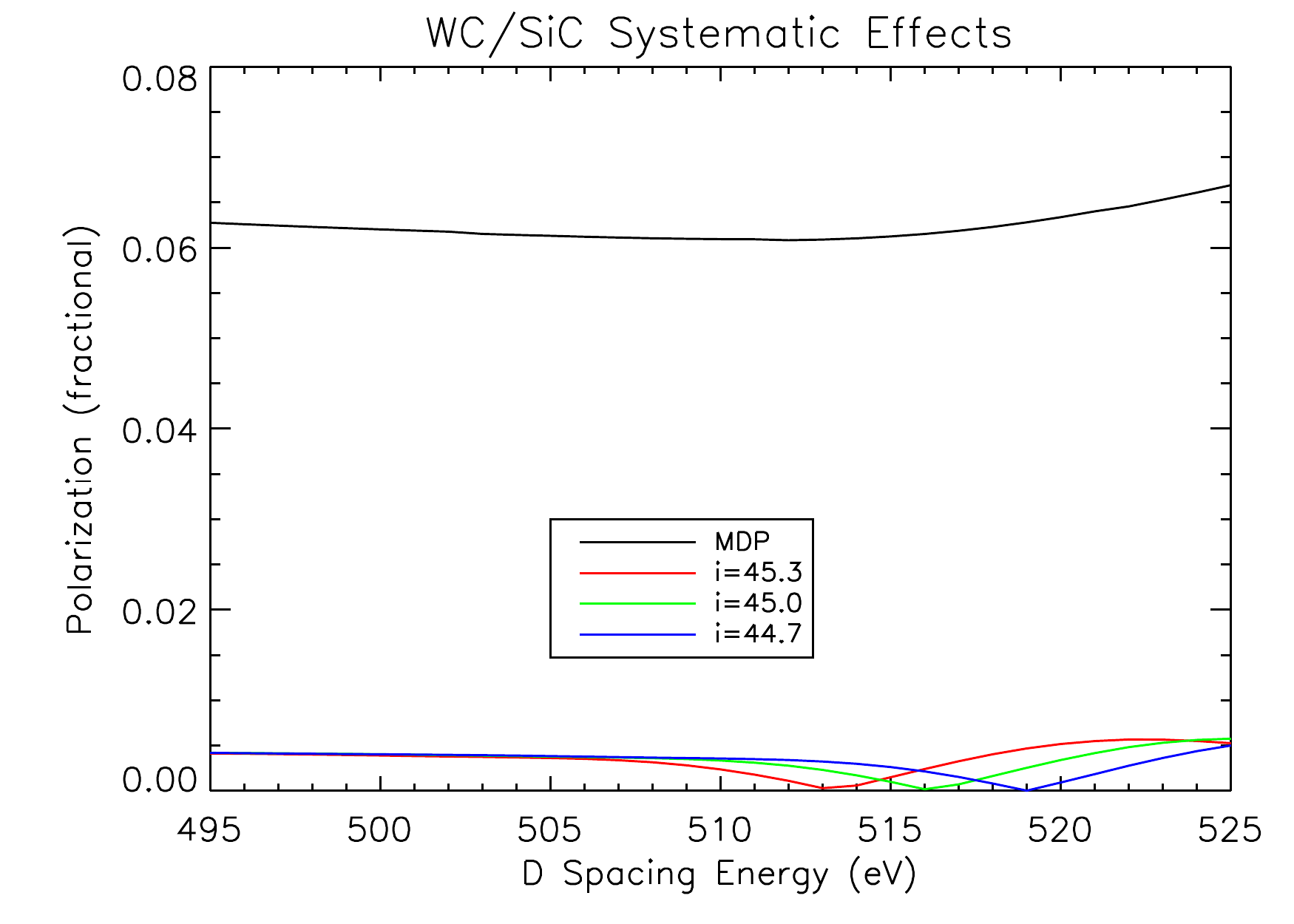}
\caption{WC/SiC false polarization predictions.  The simulation was run for zero (i=45.0), maximum (i=45.3), and minimum (i=44.7) reflector misalignment. Also plotted is optimal minimum detectable polarization for a 45$^\circ$ incidence angle from Fig.\ \ref{fig:WCOpt}.}
\label{fig:WCSystematics}
\end{minipage}
\hspace{0.5cm}
\begin{minipage}[t]{0.5\columnwidth}
\centering
\includegraphics[width=\textwidth]{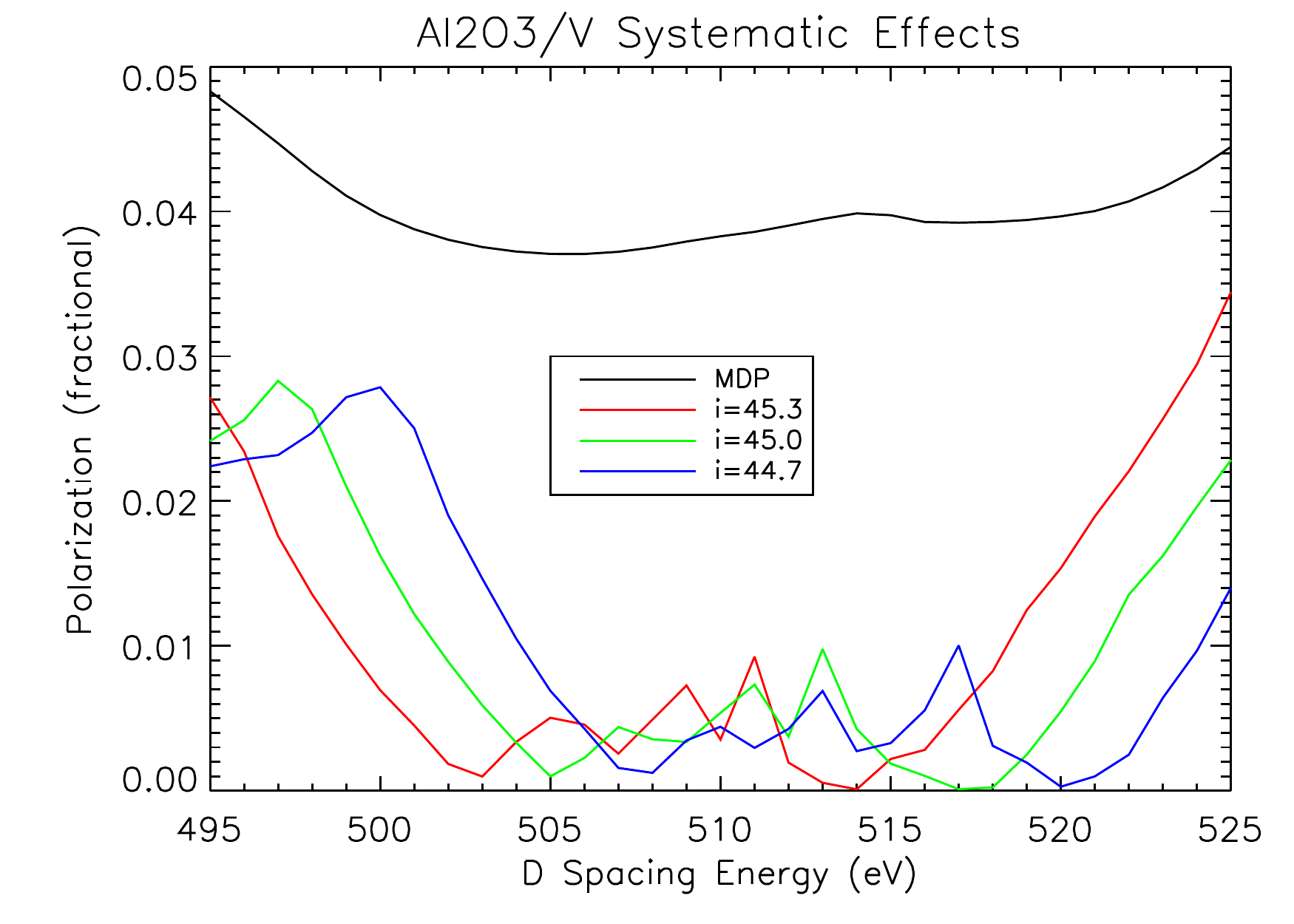}
\caption{\alo~false polarization predictions.  The simulation was run for zero (i=45.0), maximum (i=45.3), and minimum (i=44.7) reflector misalignment. Also plotted is the optimal minimum detectable polarization from Fig.\ \ref{fig:AlOpt}.}
\label{fig:AlSystematics}
\end{minipage}
\end{figure}

The results are shown in Figs.\ \ref{fig:WCSystematics} and \ref{fig:AlSystematics}.  Fractional induced polarization is plotted along with MDP against the peak reflectance energy for which the multilayer is tuned.  Although these false polarizations are certainly overestimates, it is clear the \alo~material pair has the potential to add significant uncertainty to BRP polarization measurements, particularly given that the astrophysics goal is to measure polarizations on the order of a few percent.  This effect can be attributed to the rapidly varying optical constants of vanadium near the 512 eV L$_{2,3}$ edge.

\subsection{BRP Multilayer Design}
\label{sec:MultilayerDesign}

Due to the aforementioned systematic error concerns, WC/SiC was chosen as the BRP material pair.  The MDP achieved using this material pair is slowly varying near 500 eV, so the choice of peak reflectance energy was dominated by systematic error predictions.  A peak reflectance energy of $515\pm5$ eV was chosen in order to minimize potential false polarization signals.  Using the refraction-corrected Bragg equation \cite{Spiller94}, this leads to a 1.714 nm bilayer thickness for a nominal $45^\circ$ incidence angle.  Experimental results demonstrated that a $\Gamma$ value of 0.5, combined with a 1.7 nm bilayer thickness, results in a relatively low -98 MPa (compressive) stress for WC/SiC.  Low or zero stress is highly beneficial towards the integrity of multilayer-coated free-standing membranes such as the BRP element discussed in this paper, so $\Gamma$ was fixed to 0.5.  The predicted MDP for this design is 6.1\% for $3.8\times10^5$ seconds of observation of the Crab Nebula, with $39.0\%$ margin relative to the 10\% requirement.

\section{Prototype Performance}
\label{sec:performance}

\subsection{Multilayer Deposition}
\label{sec:Deposition}

The WC/SiC multilayer coatings were deposited at Lawrence Livermore National Laboratory (LLNL) in a planar DC-magnetron sputtering system with
a ``sputter up'' geometry, where the sputtering targets (sources) are facing up and the substrate is facing down. An algorithm based on modulation of the rotational velocity of the substrate platter is used to control the coating thickness and to achieve the required coating thickness uniformity.  Each sputtering source has dimensions of 5$\times$10 in$^2$. The WC and SiC sources were operated at a constant power of 100 W and 200 W, respectively. Base pressure was maintained at $1.5-2.5\times10^{-7}$ Torr and the Ar process gas pressure was 2 mTorr.

\subsection{Reflectance and Transmission Measurements}

A prototype reflector (designated M1-110525) was fabricated on May 25, 2011, by depositing a WC/SiC multilayer coating on a polyimide substrate with a 28.6 mm aperture provided by Luxel.  The reflectance and transmission properties of the prototype reflector were measured at Beamline 6.3.2 at the Lawrence Berkeley National Laboratory Advanced Light Source (ALS) synchrotron \cite{Underwood98}.  For the transmission measurements, 600 lines/mm and 1200 lines/mm monochromator gratings were used to access the photon energy ranges 180--284 eV and 440--1304 eV, respectively.  The monochromator exit slit was set to a width of 40 $\mu$m.  Photon energy calibration was based on the absorption edges of two transmission filters (Ti, Cr) with a relative accuracy of 0.011\% rms and with 0.007\% repeatability.  Second harmonic and stray light suppression was achieved with a series of filters (C, Cr, Co, Cu, Mg). When suppression of higher-order harmonics was needed, an ``order suppressor" consisting of three Ni mirrors at a variable grazing incidence angle (depending on photon energy range) and based on the principle of total external reflection was used in addition to the filters. The ALS storage ring current was used to normalize the signal against the storage ring current decay.  The base pressure in the measurement chamber was in the range $10^{-6}$--$10^{-7}$ Torr.  The signal was collected on a GaAsP photodiode detector with a 1$^\circ$ acceptance angle. The reflectance results in the 490--550 eV region were obtained with the 1200 lines/mm grating, a Cr filter for second harmonic suppression, the order suppressor at a 6$^\circ$ incidence angle, and the GaAsP photodiode detector.

\begin{figure}[t]
\begin{minipage}[t]{0.5\columnwidth}
\centering
\includegraphics[width=\textwidth]{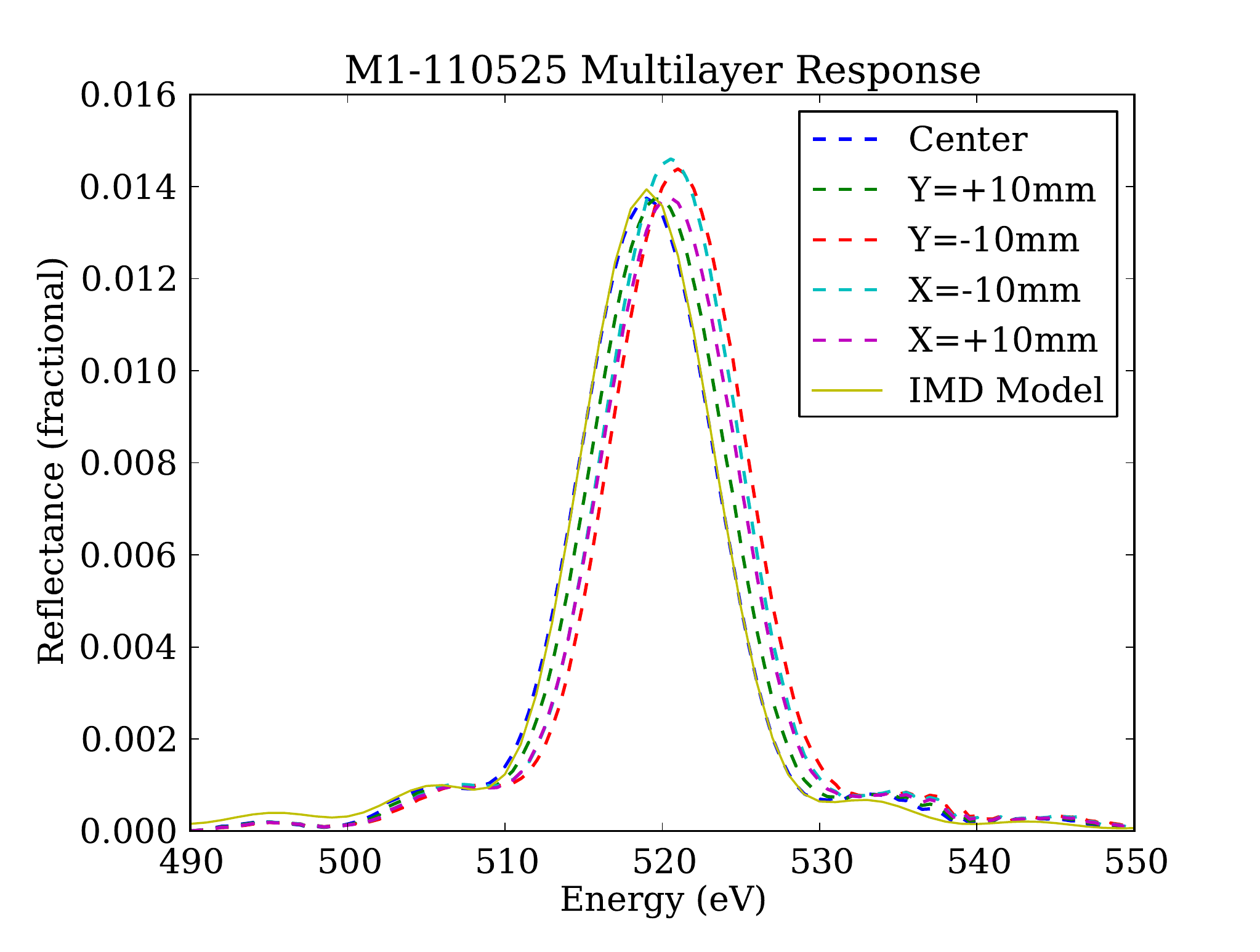}
\caption{Reflectance curves of the prototype reflector taken on June 6th, 2011.}
\label{fig:PositionPlot}
\end{minipage}
\hspace{0.5cm}
\begin{minipage}[t]{0.5\columnwidth}
\centering
\includegraphics[width=\textwidth]{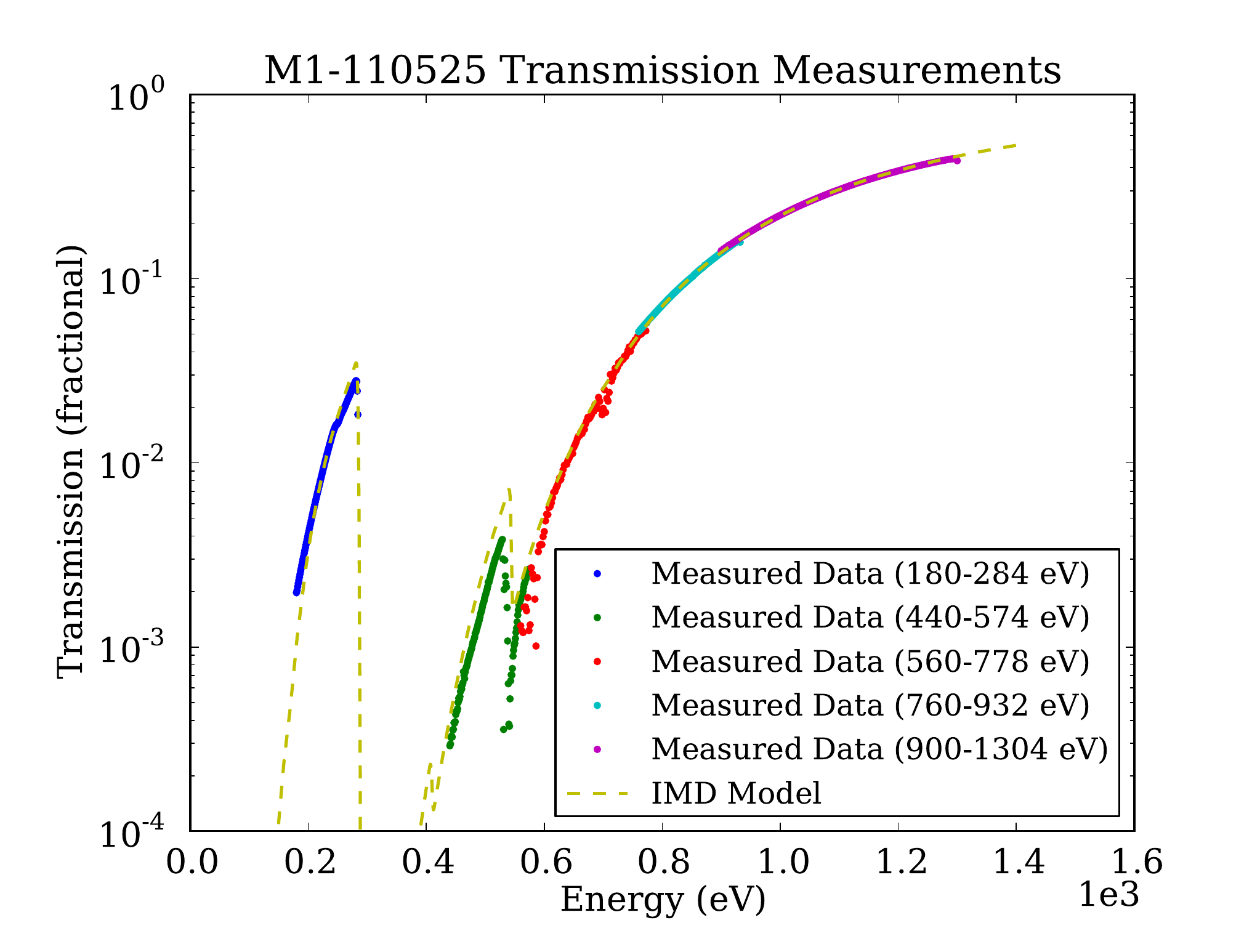}
\caption{Transmission measurements of the prototype reflector taken on June 10th, 2011.}
\label{fig:TransPlot}
\end{minipage}
\end{figure}

The first set of measurements took place June 6th, 2011.  The prototype underwent thermal soaks at the Berkeley Space Sciences Laboratory between June 8th and June 10th.  There were two $\sim 8$ hour soaks at both 0 and 40 C in a dry nitrogen environment.  The reflectance and transmission measurements were repeated on June 10th to verify the thermal stability of the prototype reflector.  The prototype underwent vibration and acoustic testing at GSFC on August 1st and 2nd.  The NASA General Environmental Verification Standard 14.1 G rms vibration profile and a 137.86 dB Pegasus class acoustic profile were used \cite{GEVS}.  Further thermal testing took place at UI in November 2011 during tests on temperature dependent ripples (see \S \ref{sec:ripples}).  Finally, the reflectance and transmission properties of the prototype were verified again at the ALS synchrotron on December 18th, 2011.

Reflectance curves from the June 2011 measurements are shown in Fig \ref{fig:PositionPlot}.  Curves taken from various points on the reflector surface (center and 10 mm off-center in four directions) indicate good bilayer thickness uniformity.  The mass density of the WC layers (15.8 g/cm$^3$) and of the SiC layers (2.98 g/cm$^3$), obtained from recent Rutherford backscattering measurements on WC and SiC sputtered thin films, were used as input parameters in IMD fits to these data \cite{Soufli09,Perea12}.  The IMD fits to these reflectance curves yielded 0.28 nm rms roughness, in good agreement with previous literature on WC/SiC coatings \cite{Jensen05}.  A 1.700 nm bilayer thickness was inferred from the IMD fits, resulting in a 519 eV peak reflectance energy as measured at the center of the sample.  The mean integrated reflectance is 0.161 eV, twice that of our requirement.  There was no observable difference between the first two sets of reflectance and transmission measurements in June---the thermal cycling at SSL had no effect on the prototype's performance.  The mean integrated reflectance, averaged over various positions on the reflector surface, measured during December 2011 was 5.6\% lower (relative) than that measured during June 2011.  Due to exposure of the reflector to non-cleanroom environments and a variety of handling conditions during the environmental testing, the reflectance reduction may be due to aging, contamination, or a combination of both.

   
The prototype's transmission was measured at the center of the sample in the energy ranges 180--284 eV and 440--1304 eV on June 10th.  The results are shown in Fig.\ \ref{fig:TransPlot}.  The yellow line is the theoretically predicted transmission based on IMD fits to the data, and matches the experimental data for the center position very closely.  Using multilayer parameters constrained by the reflectance measurements, the substrate thickness was fit to a value of 1.98 $\mu$m.  The discrepancies between the data and the theory near absorption edges are a result of the fact that the optical constants used in the IMD calculations do not include near-edge fine structure in the reflector materials.  The transmission between 440--574 eV was measured again on December 18th, and was consistent with the measurements obtained in June.  Using the IMD model, the predicted reflector transmission at 2.7 keV is 73.5\%, meeting our 70\% requirement.  Note that this result differs from our 75\% target transmission due to underestimated WC density in \S \ref{sec:modeling}.


\subsection{Ripples}
\label{sec:ripples}

      \begin{figure}
   \begin{center}
   \begin{tabular}{c}
   \includegraphics[width=\columnwidth]{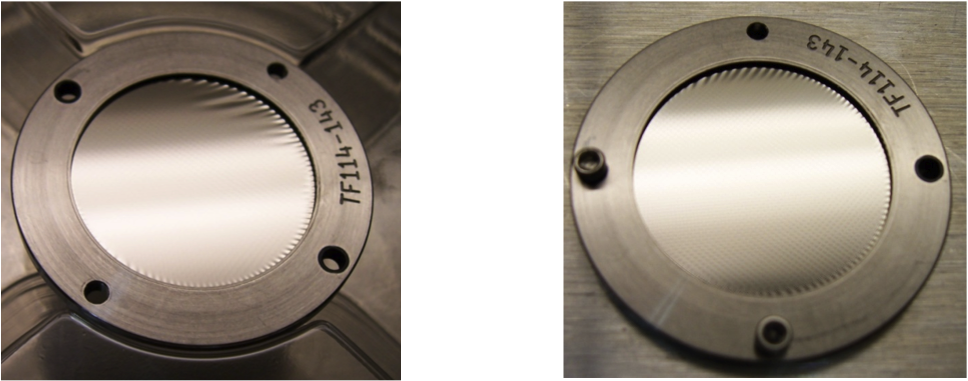}
   \end{tabular}
   \end{center}
   \caption[example] 
   { \label{fig:Ripples} 
The prototype reflector before the second set of reflectance measurements is shown at left.  At right, the prototype is shown immediately after the measurements.  The ripples receded to the perimeter of the film after the prototype was exposed to vacuum for several minutes.}
   \end{figure} 

After multilayer deposition on the polyimide substrate, ripples were noticed over an extended region of the film near the frame (see Fig.\ \ref{fig:Ripples}).  The ripples were observed to originate where regions of uneven tension existed previously on the uncoated membrane (see the ripples near the film edges in Fig.\ \ref{fig:LuxelScan}).  The ripples mostly disappeared after the coated membrane was exposed to either vacuum or dry nitrogen for several minutes.  In fact, the ripples can be eliminated by holding the prototype under a clean hood for 2--3 minutes, although they return shortly after removal of the prototype from the clean hood.  It was concluded that humidity absorption/desorption from the polyimide substrate was primarily responsible for this behavior.  Due to the dry environment in low earth orbit, the humidity dependence was not cause for concern.
   
The prototype substrate was mounted on an aluminum frame with a coefficient of thermal expansion (CTE) on the order of $\sim 23$ ppm/C, while the estimated CTE of the Luxel polyimide is $\sim 5$ ppm/C.  To investigate the effect of CTE mismatch on the ripples, the prototype was installed in a vacuum chamber with an optical viewport and pumped down to $< 1$ Torr in order to deconvolve the humidity and thermal ripple dependencies.  The chamber was cycled between 0 C and 50 C in a water bath.  At 0 C, the CTE mismatch resulted in exaggerated ripples covering the entire film surface.  At 50 C, the ripples at the perimeter of the film were seen to recede even closer to the frame material.  Evidently, the higher CTE of the frame caused the film to lose tension at low temperatures, and increased the tension at high temperatures.  For the flight model reflector, three frame material options were identified to attempt to resolve this issue: Invar, Kovar, and titanium, with estimated CTEs of 2 ppm/C, 5 ppm/C, and 8 ppm/C, respectively.

\subsection{Second Prototype}

\begin{figure}[t]
   \begin{center}
   \begin{tabular}{c}
   \includegraphics[width=\columnwidth]{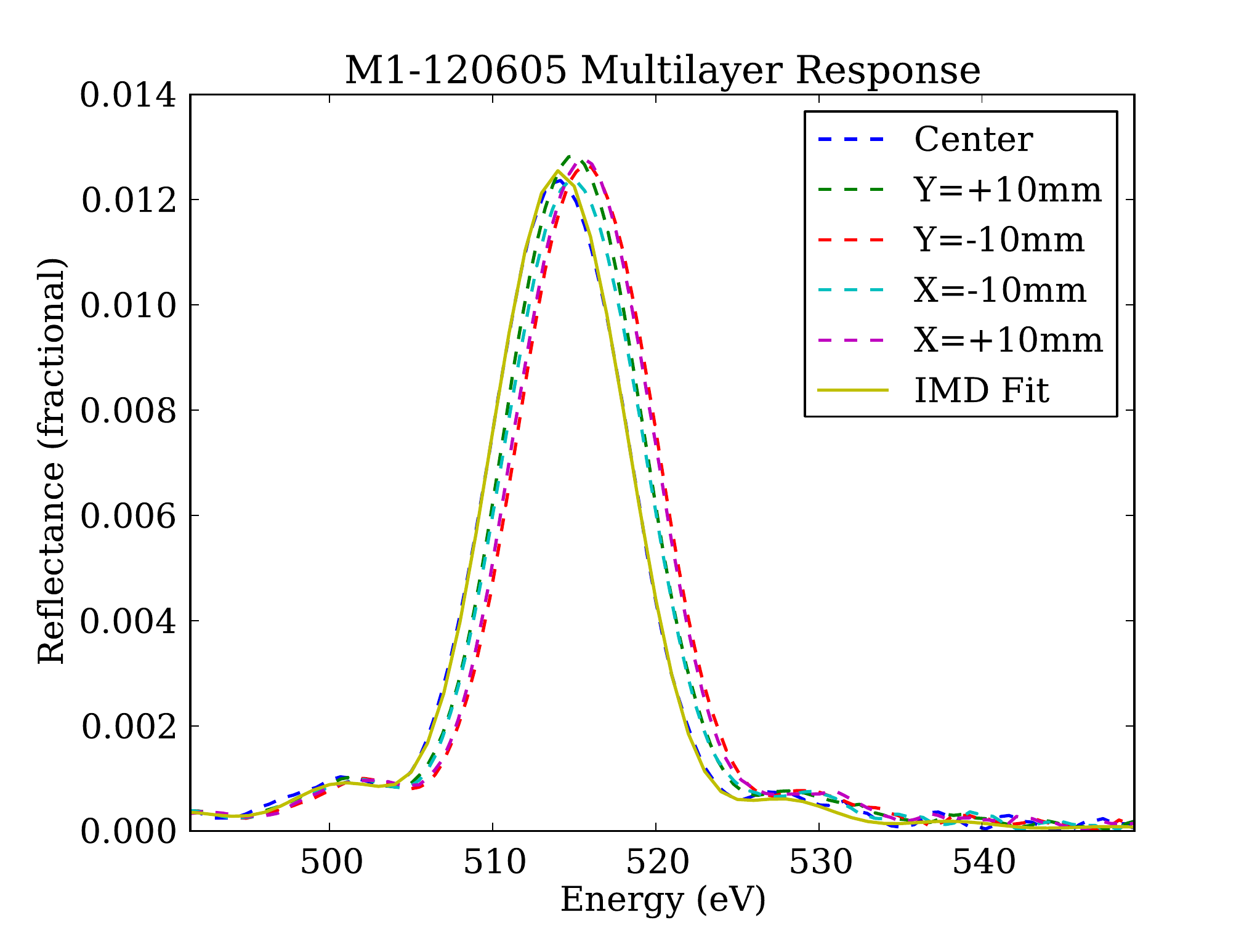}
   \end{tabular}
   \end{center}
   \caption[example] 
   { \label{fig:M1-120605} 
Reflectance curves of the second prototype reflector taken on June 14th, 2012.}
   \end{figure}

A second prototype reflector (designated M1-120605) was fabricated at LLNL on June 5, 2012 as in \S \ref{sec:Deposition}.  The target peak reflectance energy was changed to $511\pm5$ eV to facilitate a prototype BRP calibration using the vanadium 511.3 eV L$\alpha$ line, requiring a 1.728 nm bilayer thickness.  Reflectance and transmission measurements were made of this prototype at the ALS Beamline 6.3.2 on June 14, 2012.  Once again, reflectance curves were measured at five points on the reflector surface and are shown in Fig.\ \ref{fig:M1-120605}.  The average integrated reflectivity from these measurements was 0.160 eV, just below that of M1-110525.  The best fit bilayer thickness was 1.717 nm, resulting in a 514 eV peak reflectance energy as measured at the center of the sample.  The best fit substrate thickness inferred from the transmission measurements was 1.99 $\mu$m.  Extrapolating the IMD transmission model leads to a 73.3\% 2.7 keV transmission.  Ripples similar to those shown in Fig.\ \ref{fig:Ripples} were observed, again dependent on ambient humidity.

\section{Conclusions}


We have developed, fabricated, and characterized a reflective/transmissive multilayer polarizer element for operation in the 515 eV photon energy region. It consists of an ultra-short-period WC/SiC multilayer coating deposited on a 2 $\mu$m polyimide substrate of 28.6 mm diameter, mounted on an aluminum frame. Two prototype reflectors have been fabricated in this manner and meet nearly all of the GEMS BRP requirements. The reflectance and transmission achieved are more than adequate and are reliably predictable. Both reflectance and transmission properties were stable after vibration and thermal cycling tests. Ripples were observed due to thermal expansion effects and raised temperature control concerns, but selection of an appropriate frame material (Invar, Kovar, or titanium) would alleviate this issue.  Were it not for the GEMS mission cancellation, the final step would be to obtain and multilayer-coat candidate reflectors supported by frames with CTEs well-matched to that of the polyimide membrane. The reflectance and transmission, and the thermal and vibration stability of these elements would be verified using the processes discussed in this manuscript.  The reflector element with optimum performance would then be selected for flight.


\begin{acknowledgements}
The authors would like to thank Bruce Lairson of Luxel Corp.\ for providing helpful advice, many polyimide samples for analysis, and confocal microscope data. We also thank Steve McBride for use of his thermal cycling chamber, and John Tomsick for making the arrangements. Ryan Allured and Philip Kaaret are grateful to Takashi Okajima and Yang Soong at GSFC for guidance during the early stages of reflector development. We acknowledge that nearly all of our multilayer modeling was made possible with David Windt's IMD software. This work was performed under the auspices of the U.S. Department of Energy by Lawrence Livermore National Laboratory under Contract No.\ DE-AC52-07NA27344 and by the University of California Lawrence Berkeley National Laboratory under Contract No.\ DE-AC03-76F00098.  The Advanced Light Source is supported by the Director, Office of Science, Office of Basic Energy Sciences of the U.S. Department of Energy under Contract No.\ DE-AC02-05CH11231. Ryan Allured and Philip Kaaret acknowledge partial support from NASA grant NNX08AY58G.
\end{acknowledgements}



\end{document}